\documentclass[aps,prl,preprint,groupedaddress]{revtex4-2}
\usepackage{graphicx}
\usepackage{xcolor}
\usepackage[normalem]{ulem}
\begin{document}

% Use the \preprint command to place your local institutional report
% number in the upper righthand corner of the title page in preprint mode.
% Multiple \preprint commands are allowed.
% Use the 'preprintnumbers' class option to override journal defaults
% to display numbers if necessary
%\preprint{}

%Title of paper
\title{Climate network characterization of the AMOC edge state}
% repeat the \author .. \affiliation  etc. as needed
% \email, \thanks, \homepage, \altaffiliation all apply to the current
% author. Explanatory text should go in the []'s, actual e-mail
% address or url should go in the {}'s for \email and \homepage.
% Please use the appropriate macro foreach each type of information

% \affiliation command applies to all authors since the last
% \affiliation command. The \affiliation command should follow the
% other information
% \affiliation can be followed by \email, \homepage, \thanks as well.
\author{Laure Moinat}
\affiliation{Group of Applied Physics and Institute for Environmental Sciences, University of Geneva, Geneva, Switzerland}
\affiliation{Centre pour la Vie dans l'Univers (CVU), University of Geneva, Geneva, Switzerland}
\author{Reyk Börner}
\affiliation{Institute for Marine and Atmospheric research Utrecht, Department of Physics,
Utrecht University, Utrecht, The Netherlands}
\author{Valerio Lucarini}
\affiliation{School of Computing and Mathematical Sciences,
University of Leicester, Leicester, UK}
\affiliation{Schools of Sciences, Great Bay University, Dongguan, P.R. China}
\author{Maura Brunetti}
\affiliation{Group of Applied Physics and Institute for Environmental Sciences, University of Geneva, Geneva, Switzerland}
\affiliation{Centre pour la Vie dans l'Univers (CVU), University of Geneva, Geneva, Switzerland}
\author{Henk A. Dijkstra}
\affiliation{Institute for Marine and Atmospheric research Utrecht, Department of Physics,
Utrecht University, Utrecht, The Netherlands}

%Collaboration name if desired (requires use of superscriptaddress
%option in \documentclass). \noaffiliation is required (may also be
%used with the \author command).
%\collaboration can be followed by \email, \homepage, \thanks as well.
%\collaboration{}
%\noaffiliation

\date{\today}

\begin{abstract}
The Atlantic Meridional Overturning Circulation (AMOC) has been identified as a tipping element in the Earth system.
Under the current climate change scenarios, it is  urgent to develop robust methods for determining the 
probability of future AMOC transitions.   Recent studies using an Earth System Model of  Intermediate Complexity (EMIC) have 
revealed the importance of an  AMOC edge state, located on the boundary of the attraction basin of the collapsed 
state, in AMOC transitions.  Here, we  provide a characterization of this edge state through climate networks, using 
instantaneous temporal  correlations between geographical locations to define the network links.   We  apply the climate network analysis to a set of 
EMIC simulations with CO$_2$ forcing according to an intermediate climate change scenario (SSP2-4.5) that  exhibit qualitatively different AMOC  responses as a result of interaction with the edge state. We show 
that network measures, specifically the normalized degree centrality, reveal the presence of teleconnections 
across the equator as the AMOC approaches the edge state.  A similar result is obtained for an Earth System 
Model (ESM) simulating AMOC collapse or recovery, suggesting that climate networks could be used to detect 
the onset of an AMOC tipping event in ESMs. 
\end{abstract}

% insert suggested keywords - APS authors don't need to do this
%\keywords{}

%\maketitle must follow title, authors, abstract, and keywords
\maketitle

% body of paper here - Use proper section commands
% References should be done using the \cite, \ref, and \label commands
\section{Introduction}

The current climate crisis raises concerns about crossing critical thresholds of tipping elements in the Earth system~\cite{armstrong_mckay_exceeding_2022}. One of them, the Atlantic Meridional Overturning Circulation (AMOC), has been shown to play a key role in regulating the interaction between tipping elements, through its response to increased warming and freshwater input to the North Atlantic, and its impact on the heat and water cycles at the global scale~\cite{rahmstorf2024amoc,Wunderling2024}. 
There are observational indications  that the AMOC may have weakened over
the past 50 years \cite{Caesar2018,li2025amoc,Michel_2025} in addition to early warning signals of a collapse \cite{vanWesten2024}.
Moreover, climate models of different complexity have been used to investigate the effect of CO$_2$ ramping rates and freshwater hosing on the AMOC strength, showing that the AMOC can be destabilised on a centennial time scale (e.g.~\cite{liu2017,dijkstra_transitions_2026,Borner2025}). In state-of-the-art Earth system models, a transition to extremely weak overturning can occur in high-emission scenarios but also sometimes in intermediate- and low-emission scenarios~\citep{GISS_Romanou,Drijfhout_2025,oh2025noise}. Yet, many simulations also show an AMOC recovery following a transient weakening, raising the question of how to distinguish a forced transient response from the onset of an AMOC tipping event (transition to a weak or collapsed state). This calls for robust metrics on decadal time scales for AMOC destabilization. 

As deduced from the behavior of box models~\cite{Stommel1961,cessi1994,Scott1999,cimatoribus2014moc} and more complex models~\cite{marotzke1991,Lucarini2005DestabilizationTHC,rahmstorf2005MIP,Dijkstra2007,Hofmann2009,Hawkins_2011,vWestent_GRL, Manabe_Stouffer_88}, a bistable regime may exist with an ON state, which corresponds to the current state of the AMOC, and an OFF state with a weak or collapsed circulation.  Between the ON and OFF states, a so-called edge (or Melancholia) state~\cite{LucariniBodai2017} lies on the basin boundary separating the two basins of attraction, represented by the unstable branch in a bifurcation diagram. 
Although edge states are unstable, they can govern the transient dynamics of the system, and their stable sets, locally forming the basin boundary, are important for tipping behavior under time-dependent forcing.

Edge states of the climate system have been identified in a conceptual climate model composed of a chaotic atmosphere coupled to the bistable Stommel ocean~\cite{Mehling2024}, a simplified atmosphere model coupled to an ocean energy balance model (PuMA-GS)~\cite{LucariniBodai2017}, an ocean-only general circulation model (GCM)~\cite{Lohman2024}, and in PlaSim-LSG, an Earth system Model of Intermediate Complexity (EMIC)~\cite{Borner2025}. In PlaSim-LSG, the AMOC edge state is characterized by large AMOC oscillations with a characteristic centennial time scale, mainly driven by the alternating weak and strong convective regimes in the Labrador Sea~\cite{Borner2025}. 

The so-called edge tracking algorithm \cite{battelino_multiple_1988,Skufca2006} is  the main tool for identifying an edge state~\cite{LucariniBodai2017,Borner2025}. This  algorithm consists of an iterative bisection procedure that is aimed at nudging the system towards following a forward trajectory along the boundary separating the competing basins of attraction. Eventually, the system settles on the edge state, which attracts  the trajectories initialized on the basin boundary. 
In box models and an ocean-only GCM, observables exhibiting an increase in fluctuations when approaching a tipping point have been shown to be directed towards the edge state in state space~\cite{Lohmann2025}. This property could be used to identify the relevant variables showing substantial critical slowing down in a high-dimensional system~\cite{Lohmann2025}, thus revealing the potential role of the edge state to serve as an early warning signal of an AMOC instability. 

Applying the edge tracking algorithm to comprehensive Earth System Models (ESMs)  is not feasible at the moment due to computational limitations. 
Therefore, we propose an alternative  way to identify the edge state using  climate networks, which combine both spatial and temporal information of dynamical processes~\cite{Dijkstra2019NetworksClimate}. Climate networks have been successfully used to identify linear and nonlinear properties of the climate system~\cite{Donges2009, Donges_2015, Donges_2015_q}, to describe teleconnections~\cite{Strnad2022}, to investigate the dynamics of climate subsystems as the AMOC~\cite{vandermheen2013} and El Ni\~no-Southern Oscillation~\cite{radebach2013}, or global-scale tipping~\cite{moinat24}, and as a mathematical tool to identify properties that cannot be directly inferred from time series analyses~\cite{Donges2009}.  Climate network analysis has been 
used to determine early warning signals of an AMOC collapse in the FAMOUS 
model~\cite{Hawkins_2011,Feng2014}, with 
the kurtosis of the degree distribution as main indicator. 

From a dynamical systems point of view, the onset of an AMOC collapse may be defined as the trajectory exiting the (forcing-dependent) basin of attraction of the AMOC-ON state when crossing its basin boundary. Furthermore, the AMOC-ON state may lose stability by colliding with the edge state in a so-called boundary crisis marking a critical forcing threshold~\cite{Borner2025}. This highlights the central role of the edge state for AMOC tipping. While transitions under transient forcing are not guaranteed to travel through the edge state, they are likely to interact with its dynamics especially under near-critical forcing. The recovery of the AMOC following a substantial excursion towards 
the OFF-state can also be associated with a close approach to the edge state. Hence, we expect to be able to identify the 
edge state's spatial signature in climate model simulations under different emission scenarios where the AMOC shows 
collapse or recovery. 

The paper is organized as follows. First, we briefly describe the PlaSim-LSG simulations that are used in the present study, how we construct the climate networks  \cite{Donges_2015} and the network indicators 
used to perform the analysis.  Next, to identify relevant observables, we compare the climate network constructed from the ON state in the PlaSim-LSG 
simulation under pre-industrial conditions against that constructed from reanalysis data.  
Using these observables, we construct and characterize the climate networks of the ON, OFF and 
EDGE states.  We also investigate the temporal evolution of network indicators for an initial condition ensemble of AMOC 
trajectories under an intermediate CO$_2$ emissions scenario, that were shown to explicitly go through the edge state using the tracking 
algorithm \cite[]{Borner2025}. Finally, we  apply the same network  analysis to simulations obtained by the NASA GISS model~\cite{GISS_Romanou} under the same scenario, an ESM participating in the Coupled Model Intercomparison Project (CMIP) 
phase 6. 

\section{Methods}
\subsection*{Climate model simulations}
\label{sec:sims}

The present analysis is based on the results of the climate simulations described in~\cite{Borner2025} and obtained using PlaSim-LSG~\cite{Angeloni2022}. For  pre-industrial forcing conditions where the CO$_2$ concentration is set to 285~ppm,  
three equilibrium states of the AMOC were obtained: ON, OFF and EDGE states. The ON state corresponds to the pre-industrial AMOC, the OFF state shows a weakened AMOC with a shutdown of North Atlantic deep convection, and the EDGE state has an intermediate AMOC strength with centennial-scale oscillations. 

These three states were also obtained at present-day forcing conditions (360~ppm), and we will use the simulations at 360~ppm to explicitly characterize the EDGE state behavior because they provide longer time series than at 285~ppm (see \cite{Borner2025}). Starting from the ON state at 285~ppm, novel transient PlaSim-LSG simulations were performed following the increase in CO$_2$ of the Shared Socioeconomic Pathway (SSP) 2-4.5 \cite{Meinshausen2020_SSP}, an intermediate emission scenario, for five ensemble members until the year 2500 \cite{Borner2025}.

\begin{table}[htbp]
\centering
\caption{Datasets and climate model characteristics for the different experimental configurations. For the GISS model, the ensemble members are reported explicitly.}
\label{tab:climate_datasets}
\renewcommand{\arraystretch}{1.3}
\begin{tabular}{lcccc}
\hline
 & PlaSim-LSG & ORA20C & NASA-GISS-E2-1-G & Analysis\\
\hline
Resolution       & 3.5° $\times$  3.5°          & 1° $\times$ 1°            & 1° $\times$ 1.25° & 2°$\times$2° regridded\\
Time interval [yr] & 25, 50 & 25 & 50 & All\\
piControl        & 285 ppm     & 1900 -- 1925   & {r1i1p1f2} & Observable selection\\
Present-day  & 360 ppm     & NA           & NA & State characterization \\
SSP2-4.5         & 5 members  & NA           & {r1i1p1f2}, {r1i1p1f10}, {r1i1p1f7} & Network indicators  \\
\hline
\end{tabular}
\label{table:exp}
\end{table}

In addition, we use data from several simulations with the NASA GISS-E2-1-G model~\cite{GISS_Romanou}  that have been highlighted in the qualitative comparison between the GISS and PlaSim-LSG models performed 
in Ref.~\cite{Borner2025}. In the GISS model, the ocean has a resolution of 1° $\times$ 1.25° with 40 vertical levels and was re-gridded to a resolution of 2° $\times$ 2° to limit the computational time required by the network analysis. The three extracted ensemble members are r1i1p1f2, r7i1p1f2 and r10i1p1f2 for the SSP2$-$4.5 scenario, with their corresponding historical data. The piControl simulation from r1i1p1f2 is used as the ON state reference for the three ensembles, since it is the only one available.
On all the transient simulations for PlaSim-LSG and NASA-GISS, the climate networks are performed with a moving time window of 25 (appendix) or 50 years (main text).

In order to determine the most appropriate observables  for the network construction, the ORA20C reanalysis (which combines a 10-member ensemble of reanalyses) \cite{deBoisseson2018} is used.   In ORA20C, the 1900-1925 period with an annual resolution is taken; hence we obtain a time series of 25 years. ORA20C has a resolution of 1° $\times$ 1° and was re-gridded to fit the PlaSim-LSG resolution of around 3.5° $\times$ 3.5°. 

\subsection*{Climate networks}
\label{sec:networks}

Climate networks are statistical tools to unveil the structure of interactions (links or connections) between subsystems (nodes or vertices) of a complex system. Here, we consider pair interactions between annual-mean time series at the grid points of climate model output.  In this case, the $N$ nodes of the network consist of all horizontal grid points of the ocean output. To construct the network, we consider  temperature and salinity time series at the surface (top 50~m) and at depth (vertically averaged between 300~m and 1000~m). 

The undirected and area-weighted climate networks are computed using the pyUnicorn package~\cite{Donges_2015}, and links are generated with the Pearson correlation coefficient (PCC) at a fixed threshold of 0.5 and time lag 0, using the \textit{climate.TsonisClimateNetwork} function. This means that links between two time series at different grid points are established if the correlation coefficient is larger than 0.5. This threshold has been chosen to avoid the saturation effect that occurs when the threshold is too small and, at the same time, to guarantee a sufficient number of links, as shown in Ref.~\cite{moinat24}.
To generate the networks, a window size of 25 or 50 years is used for several reasons: sufficiently long time series are needed to generate statistically robust networks; the availability of observational data; and the duration of strong and weak phases of the EDGE state oscillations. 

Correlations at zero lag capture how ocean variables co-vary instantaneously (i.e. on annual timescales) at different locations. Long-range interactions must therefore be mediated by relatively fast processes, e.g. atmospheric teleconnections. This is different from considering the advective transport of mass, energy and information between regions on longer timescales.

Our network analysis is based on the adjacency matrix and two global network measures, namely the normalized degree centrality and the average length distance, since they have been identified as relevant indicators when looking at global-scale tipping behavior~\citep{moinat24}. The adjacency matrix is an $N \times N$ matrix that contains the information on the links: 1 is assigned between a pair of grid points if the correlation is above the threshold, and 0 if not. The normalized degree centrality (ND) measures the number of links per nodes, normalized by the total number of possible links. The average length distance (ALD) measures the topological distance between two connected vertices~\cite{Dijkstra2019NetworksClimate}. 

The adjacency matrix and the ND and ALD  indicators are used to extract information about the AMOC states. 
In order to evaluate the number of links among two regions,  we extract from the adjacency matrix the vertices in 
the corresponding regions. Then, the density of connections is given by the ratio between the links of one region 
to the other and the total number of possible links between these two regions. Noise is removed by setting a 
threshold at 5\% in the connection density.

\section*{Results} 

\subsection{Relevant climate observables}

\label{sec:variable}

\begin{figure}  
\centering
\includegraphics[width=0.5\textwidth, keepaspectratio]{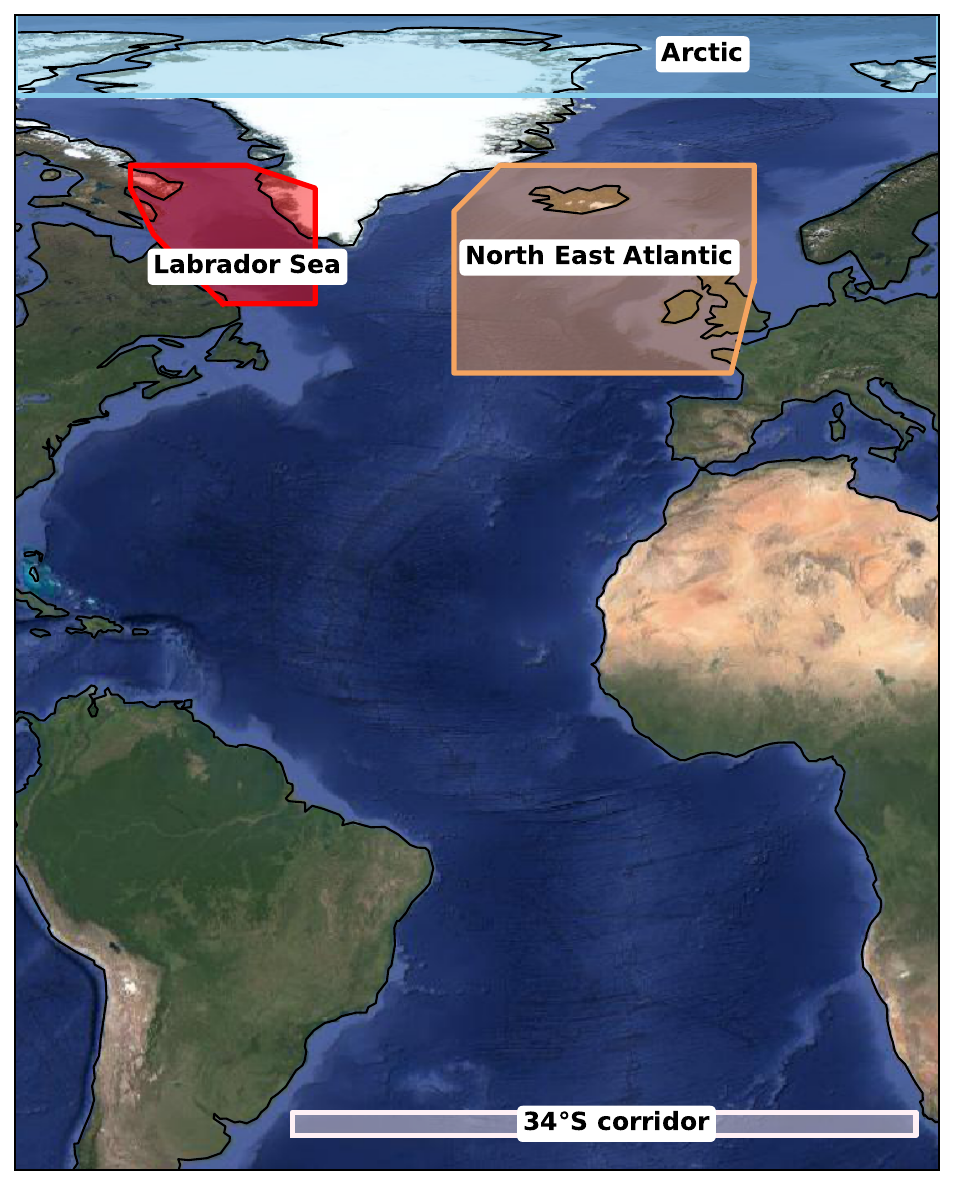}
\caption{Regions used to compute the connection density in the climate network. They include the Arctic (light blue), the Labrador Sea (red), the North-East Atlantic (light brown) and the 34°S corridor (light pink). }
\label{regions of interest}
\end{figure}

This study focuses on the North Atlantic basin and selected sub-regions to better understand the potential teleconnections
between them. Four regions are identified as relevant: the two main convective regions in the model, the Labrador Sea and the subpolar North 
East Atlantic; the Arctic Ocean, as sea ice has been identified to play an important role in the EDGE state 
behavior~\cite{Borner2025}; the Southern border of the Atlantic (34°S corridor), as it has been shown to be a key region for AMOC stability~\cite{vanWesten2024}. The exact delimitation of these regions can be found in Fig.~\ref{regions of interest}.  

To compare the climate networks from the  PlaSim-LSG ON state with those from observations,  networks were generated 
from the  ORA20C reanalysis in the 1900-1925 period and from the pre-industrial PlaSim-LSG simulation, more precisely by comparing region-to-region connections. 

For the ON state in PlaSim-LSG, the entire simulated period is considered and cut into chunks of 25 years, which are 
then averaged, due to the equilibrium nature of this simulation. This evaluation was performed for sea temperature and 
salinity at the  surface and at depth (average values between 300~m and 1000~m). The difference between the raw fields can be seen in Fig.~\ref{fig:on_validation_ORA20c}. 
\begin{figure}  
\centering
\includegraphics[width=1.0\textwidth, keepaspectratio]{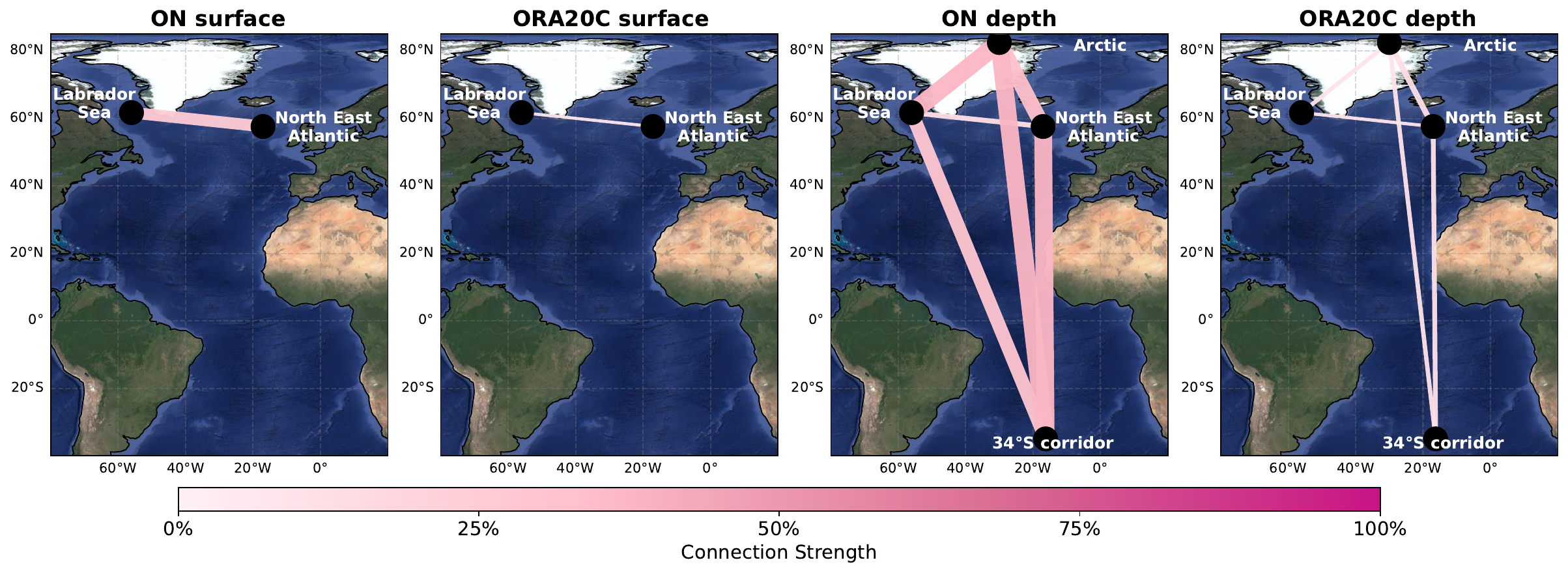}
\caption{Region-to-region temperature connections for the ON state at surface, ORA20C at surface, ON state at depth and the ORA20C at depth, respectively. The darkness of the pink color and the thickness of the connecting lines are proportional to the connection density. A 5\% threshold has been applied to remove the noise. The region labels are the ones displaying connections with other regions.}
\label{on state regions}
\end{figure} 

As shown in Fig.~\ref{on state regions}, the networks generated using sea surface temperature in PlaSim-LSG and in ORA20C display the same region-to-region connections, namely between the Labrador Sea and the North East Atlantic, with a comparable intensity.
For the networks constructed from potential temperature at depth, the intensity of the connections in the ON state is in the 30\% range, while that in ORA20C is less than 15\%. The Labrador Sea is not directly connected to the 34$^\circ$ corridor in ORA20C, while the other region-to-region connections are qualitatively the same in the two cases. 
Thus, we  conclude that the inter-regional network connections based on the ocean temperature in the PlaSim-LSG ON state at 285~ppm capture the main dynamical features of the network connections in ORA20C. This is not satisfied by the networks constructed using the salinity field, neither at the surface nor at depth. Freshwater biases are a common issue in global climate models \cite{vWesten2024_OS} and could also cause the discrepancies in the network structure of the salinity field. Additionally, PlaSim-LSG has several important limitations in representing ocean dynamics, including the coarse resolution, lack of eddy parameterization and a simplified convective adjustment scheme \cite{Angeloni2022}.

\section{State characterization}
\label{sec:StateCharacterization}

We will next use the 360~ppm Plasim-LSG ON state  simulation (see Table~\ref{table:exp}). In this case, the ON state 
acts as the reference for the network behavior. As with the ON state at 285~ppm, at the surface there are weak connections between the Labrador Sea and the North-East Atlantic, and additional weak connections between the Labrador Sea and the Arctic region. This could be related to the retreat of the sea ice extent in the Arctic region under warming conditions, possibly enabling atmospheric teleconnections to the Labrador Sea. At depth, we see the appearance of the connections of the 34°S corridor with the Arctic region, the North-East Atlantic and the Labrador Sea, as in the 285~ppm case.

The OFF state displays a behavior that is different from the ON state. At the surface, only the Labrador Sea and the North-East Atlantic are connected, while the Arctic region does not connect with these two regions. 
In this case the Arctic region is fully covered by sea ice \cite{Borner2025}, since, because of the presence of a weak AMOC, the sea ice has expanded southwards. As a result, the surface Arctic Ocean may be effectively decoupled even from the neighboring regions.  
At depth, there is an overall decrease in intensity of the connections between the northern and southern hemispheres, while the coupling increases within the northern hemisphere. This can be linked to the collapsed overturning north of 50$^\circ$N and associated shutdown of deep convection, likely affecting the connectivity of the overturning circulation across latitudes.

\begin{figure}  
\centering
\includegraphics[width=1.0\textwidth, keepaspectratio]{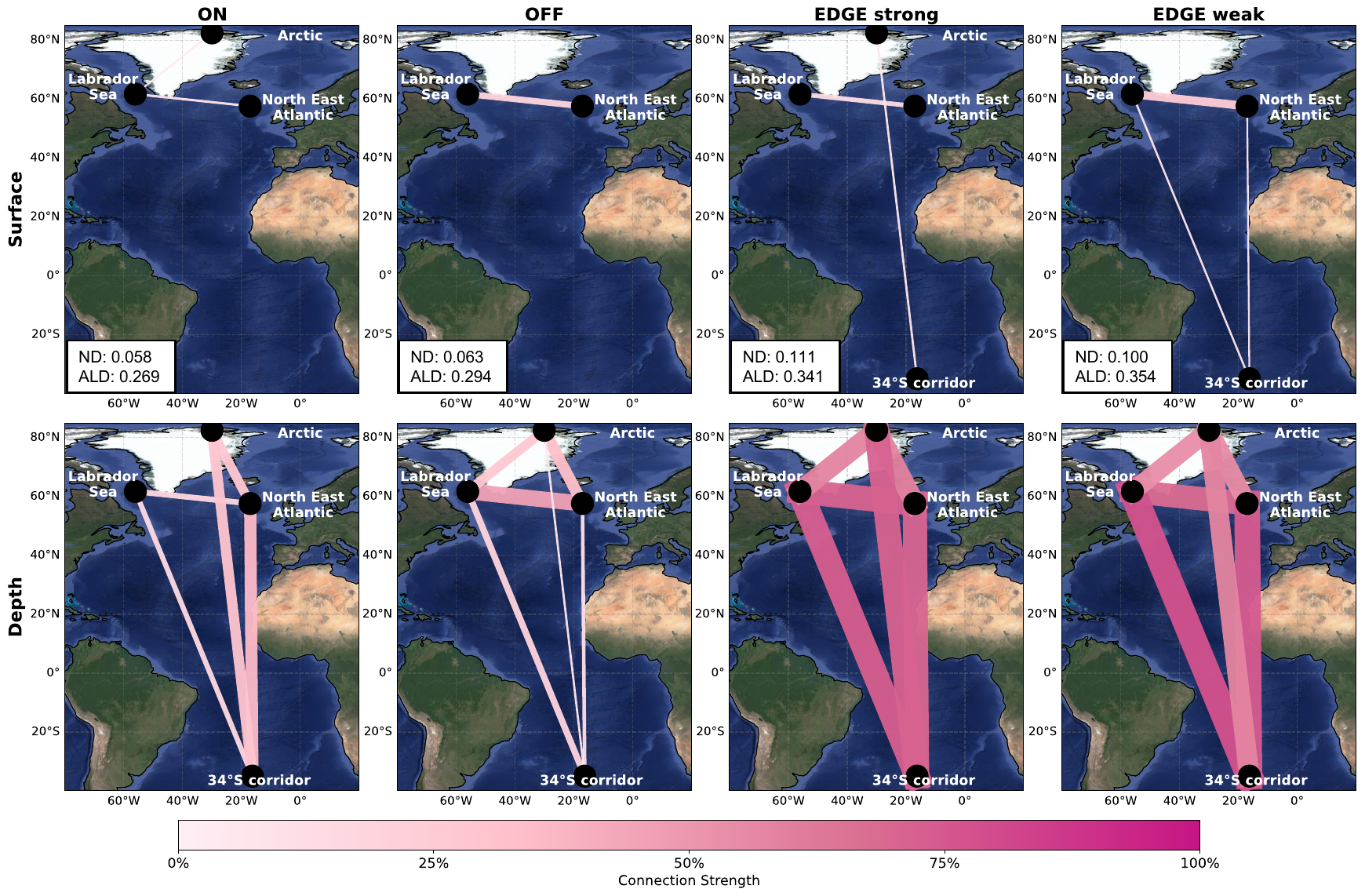}
\caption{Region-to-region connections for the ON (first column), OFF (second column), EDGE strong (third column) and weak (fourth column) states at the surface (first line) and at depth (second line) obtained under the 360 ppm CO$_2$ value. For the surface, the average values of the normalized degree centrality (ND) and the average length distance (ALD) of the Atlantic basin are also shown in the inset. 
The darkness of the pink color and the thickness of the connecting lines are proportional to the connection density. A 5\% threshold has been applied to remove the noise. The region labels are the ones displaying connections with other regions.}
\label{fig:EDGE-weak-strong-regions}
\end{figure}

\begin{figure}  
\centering
\includegraphics[width=1.0\textwidth, keepaspectratio]{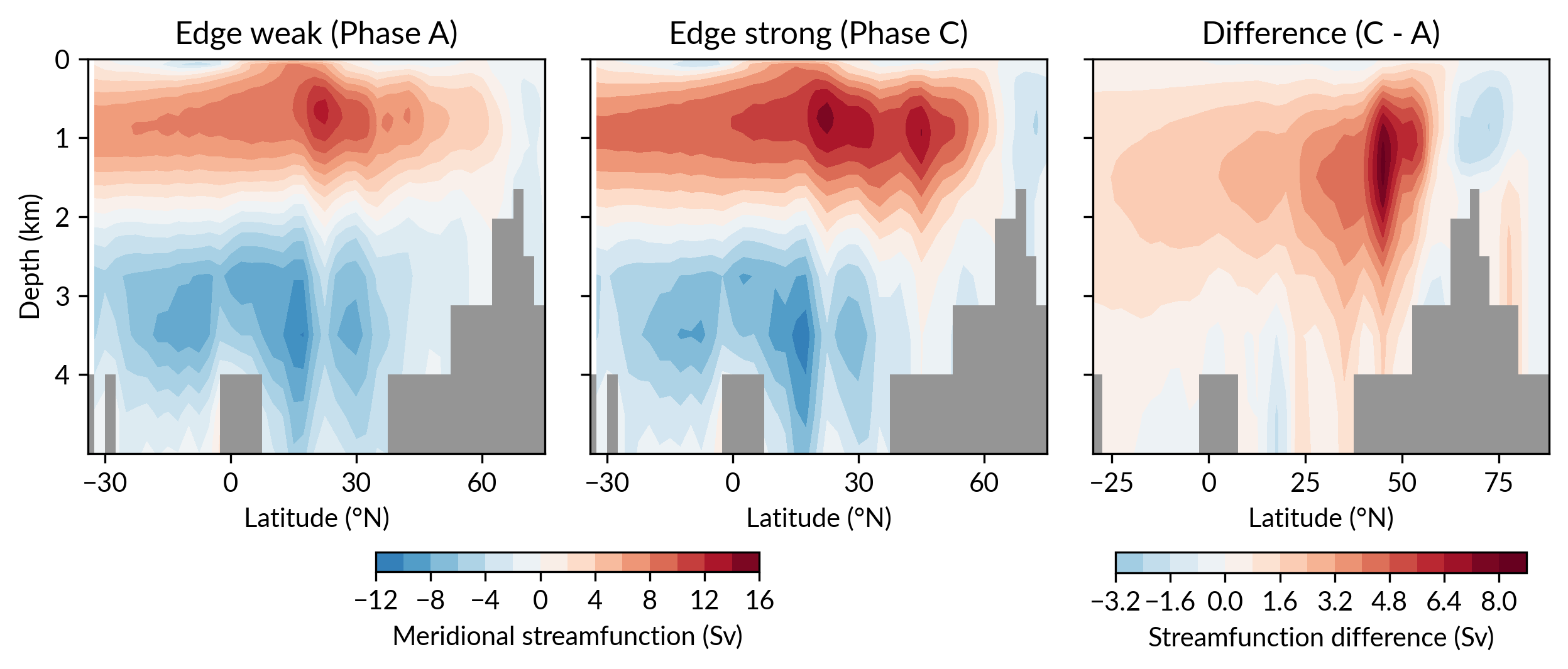}
\caption{Overturning streamfunction in the Atlantic basin for the EDGE weak/phase \textit{A} (first panel), for the EDGE strong/phase \textit{C} (second panel) and the difference between the EDGE strong and weak (third panel) with the corresponding colorscales.}
\label{fig:AMOC_EDGE}
\end{figure}

Regarding the EDGE state, the networks have been computed separately for the weak (\textit{A}) and strong (\textit{C}) phases of the oscillation 
identified in \cite{Borner2025}, corresponding to weak and strong convection in the Labrador Sea, respectively (Fig.~\ref{fig:AMOC_EDGE}). 

At the surface, we see that teleconnections  with the 34°S corridor appear for both weak and strong convection 
regimes, hence unveiling that, in the unstable EDGE state, teleconnections occur across the Equator. 
At depth, the connections among all regions are substantially intensified as compared to the ON and OFF states. 
These connections occurring on decadal time scales between the northern and southern hemispheres have been 
seen in simpler models, where following a perturbation both hemispheres react coherently and change 
together \cite[]{Scott1999, Lucarini2005b, Lucarini2005DestabilizationTHC}. They are also consistent with the 
complex network analysis of an AMOC collapse in the FAMOUS model, where high degree nodes appear 
near the transition over the whole depth \cite[]{Feng2014}. 

The salinity and temperature fields during the weak and strong convective regimes of the EDGE state 
are shown in Fig.~\ref{fig:differenceT}. 

 In both convection regimes, the subpolar North Atlantic surface ocean is cooler and fresher compared to the ON state, as less heat and salt is advected 
 from the tropics, while at the same time a slightly warmer and saltier region emerges at the 34°S corridor, in 
 particular on the western boundary of the South Atlantic Ocean. At depth, the subtropical and tropical Atlantic is substantially warmer compared to the ON state.

To quantify the difference between the strong and weak phases of the EDGE state, we calculated the correlation between the normalized degree centrality in each network, which are $R =0.19, p=0.002$ for the surface and $R = 0.47, p<0.001$ at depth. For both depths, the correlation coefficient is positive and significant. However, the $R$ value at the surface remains low, emphasizing the need to distinguish between the convective regimes in the EDGE state. This is also seen in the different patterns of region-to-region connections in the North Atlantic basin at the surface (Fig.~\ref{fig:EDGE-weak-strong-regions}). 

At the surface and at depth, the EDGE state has a characteristic network signature, with significantly more intense connections with respect to the ON and OFF states, and the appearance of teleconnections with the Southern Hemisphere. Indeed, during a tipping process, the number of spatial correlations increases, similarly to what is observed in 
thermodynamic phase transitions~\cite{Dijkstra2019NetworksClimate}. 
This behavior can be understood through the so-called slaving principle, according to which, near a bifurcation, all dynamical modes rapidly adjust to the slowly evolving dominant mode, leading to the emergence of spatial correlations that extend across the entire system~\cite{Dakos2010SpatialCorrelation,Dijkstra2019NetworksClimate,moinat24}. In many networked systems this is accompanied by critical slowing down and emergence of synchronization \cite{ZagliLucariniPavliotis,ZaglietalJPA2024}. We may thus interpret the emergence of interhemispheric teleconnections as a manifestation of a characteristic network signature of an edge state.

\section{Proximity to the EDGE state}
\label{sec:signature}

\begin{figure}  
\centering
\includegraphics[width=1.0\textwidth, keepaspectratio]{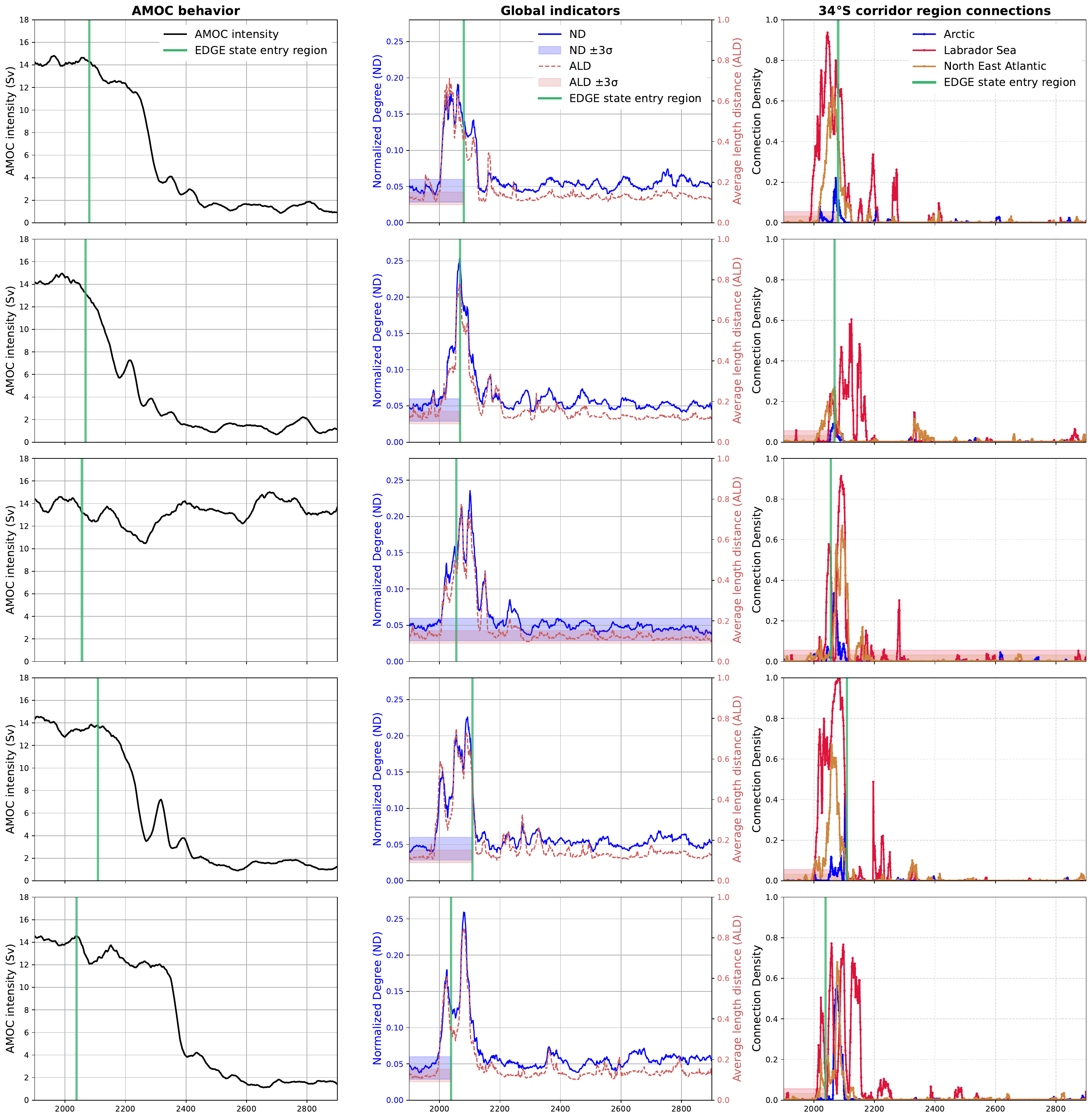}
\caption{Temporal evolution of the 30-years moving average AMOC intensity (first column), global network indicators (second column) and  region-to-region connections starting from the 34°S corridor (third column) in five PlaSim-LSG simulation ensemble members. The green vertical line corresponds to the time when the AMOC enters the EDGE-state region, as inferred from the reduced salinity phase space in~\cite{Borner2025}. }
\label{PLASIMsurface}
\end{figure} 

\begin{figure}  
\centering
\includegraphics[width=1.0\textwidth, keepaspectratio]{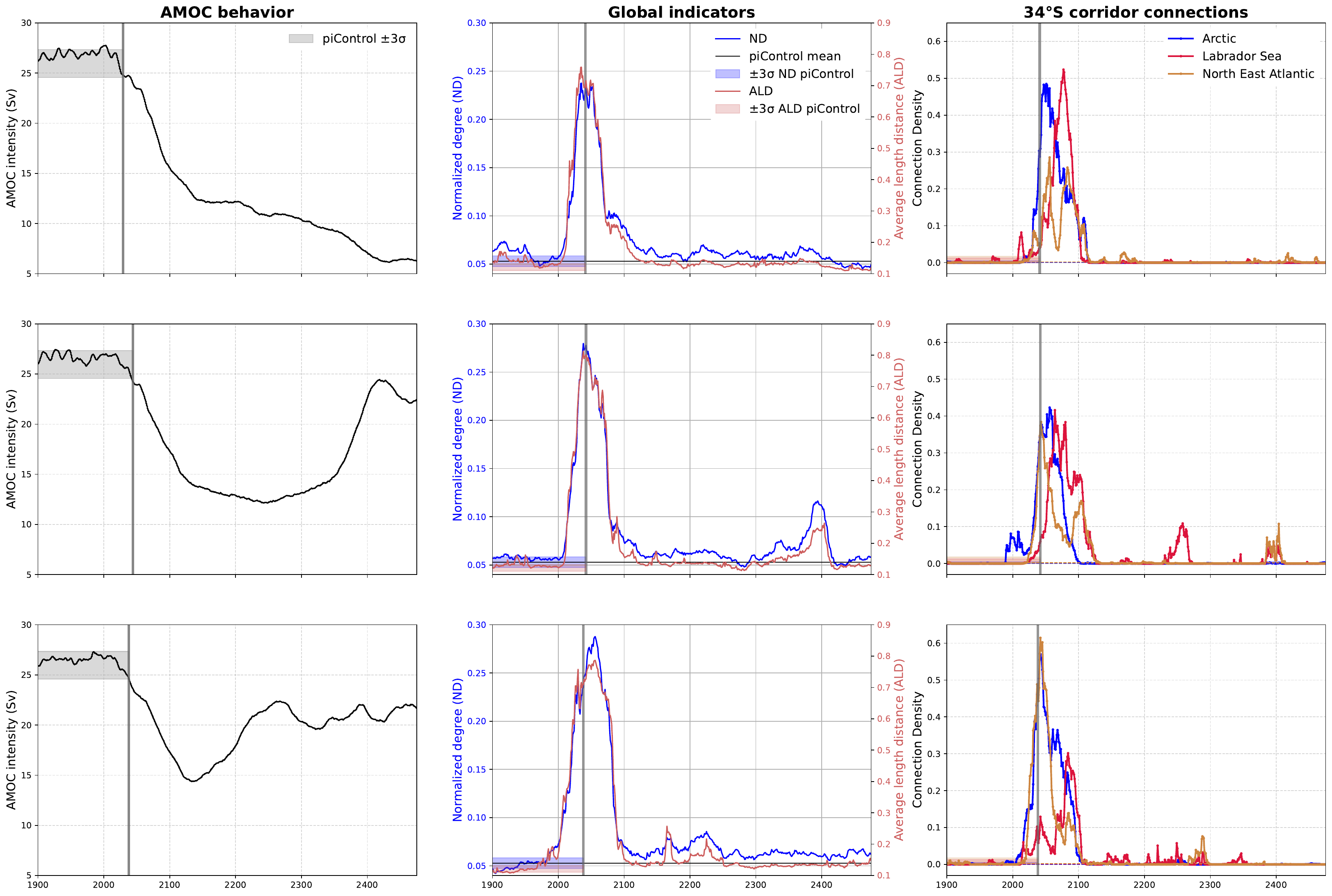}
\caption{Same as Fig.~\ref{PLASIMsurface} in three GISS simulation ensembles (r1i1p1f2, r7i1p1f2 and r10i1p1f2, respectively). The gray vertical line corresponds to the time when the AMOC strength exits the 3$\sigma$ band (shaded area).}
\label{GISSsurface}
\end{figure} 

We now apply the network analysis to simulations under time-dependent CO$_2$ forcing representing the SSP2-4.5 scenario. The goal is to use the identified network signature of the EDGE state to detect the approach to the EDGE state in the transient simulations.
The AMOC intensity is calculated as the streamfunction maximum in the 46 - 66°N latitude range (PlaSim-LSG) and the maximum value at 26.5°N below 500 m (GISS), consistent with the conventions in Refs.~\cite{Borner2025} and \cite{GISS_Romanou}, respectively. 
Figure~\ref{PLASIMsurface} displays the evolution of the AMOC intensity (first column), the global ND and ALD (second column), and region-to-region connections of the 34$^\circ$S corridor with the other three regions defined in Fig.~\ref{regions of interest}, for the 5-member ensemble in PlaSim-LSG. 
All ensemble trajectories lead to the EDGE state region under the transient forcing, but only four members undergo a subsequent AMOC collapse while one member maintains a relatively strong AMOC.

To evaluate the variability in the ON state, the mean value of each indicator and each region-to-region connection density is computed with its standard deviation $\sigma$. Then, a 3$\sigma$ band is represented as shaded areas in the different subplots.  A moving window of 50 years is used. Using a 25-year moving window produces similar but noisier results (Fig.~\ref{fig:amocbehaviour25yrs}). For comparison, we also perform the analysis for temperature at depth (Fig.~\ref{fig:50yrPlasimDepth}, using again a time window of 50 yr), but we focus on the surface in the following since the signal is most clear there. 

In Fig.~\ref{PLASIMsurface}, a green vertical line indicates the time when the AMOC state enters the EDGE-state region, as inferred from the reduced salinity phase space in~\citep{Borner2025}. The 34°S corridor shows connections to other regions before entering the EDGE state region, regardless of whether the eventual fate of the AMOC is collapse or recovery. 
The global indicators (namely, the normalized degree centrality and the average path length in the second column of Fig.~\ref{PLASIMsurface}), exit the 3$\sigma$ variability band well before entering the EDGE state region,  suggesting that the system is exploring an unusual dynamical regime. The key point here is that the proximity to the EDGE state (rather than the OFF state) is rather well represented, 
as can be seen by comparing the results  obtained for both the collapsed and recovery AMOC cases; the latter shown in the third row of Fig.~\ref{PLASIMsurface}. This result provides a new method to unveil the presence of an EDGE state in complex climate models where it is currently too computationally expensive to explicitly compute it.

The selected GISS simulations comprise one collapsing member, one with late recovery and one with early recovery (Fig.~\ref{GISSsurface}; see also Fig. 14 in~\cite{Borner2025}). In the GISS model, the presence and location of an AMOC edge state is not known, but the divergence of ensemble members suggests the probabilistic crossing of a basin boundary. The normalized degree centrality (blue lines in the second-column panels) and the region-to-region connections from the 34°S corridor (third-column panels) are good indicators of AMOC weakening (corresponding to the vertical gray line in the first-column panels, where the AMOC intensity becomes lower than the 3$\sigma$ band of AMOC variability in the ON state), both at the surface (Fig.~\ref{GISSsurface}) and at depth (Fig.~\ref{fig:GISSdepth}), albeit the signal is clearer at the surface. 
The normalized degree centrality and connections between 34°S corridor and Labrador Sea also seem  indicators for the recovery phase (around year 2300 in the central-row panels).

It is important to note that to obtain a clear Early Warming Signal (EWS), several indicators should display a peak above the 3$\sigma$ band at the same time, while peaks in only one indicator may correspond to false positives.  The agreement between the evolution of the normalized degree centrality in the North Atlantic basin and the region-to-region connections provides a coherent picture showing that the system is going away from the ON state, suggesting that all members get close to the basin boundary. This coherence is also seen at depth (Fig.~\ref{fig:GISSdepth}), despite being delayed. In addition, the similar  results between the  Plasim-LSG  and GISS models suggest that 
the GISS model comes close to an EDGE state as it has been suggested in \cite{Borner2025}.  

The 34°S corridor teleconnections behave as an important indicator for moving away of the ON attractor and towards the EDGE state in all cases. 
As identified earlier in the characterization of the states (Fig.~\ref{fig:EDGE-weak-strong-regions}), the EDGE state is characterized by the appearance of these interhemispheric teleconnections. 
In all cases, 34°S corridor teleconnections emerge around the year 2000. The signal at depth confirms the peaks in teleconnections for the EDGE state across the Equator (Fig.~\ref{fig:50yrPlasimDepth}). This conclusion can also be applied to the GISS runs, hence emphasizing that the EDGE state behavior does not seem to be unique to PlaSim-LSG. This confirmation by a comprehensive ESM suggests that the network signature could be robust across models of varying complexity.

The order-of-magnitude amplification of interhemispheric teleconnections at the beginning of the 21st century, seen in both the PlaSim-LSG and GISS models, unveils the promising role of spatial correlations in the search for EWSs \cite{lenton2024remotely}. Nevertheless, to minimize the number of false positives, several indicators should display the same behavior to have a robust EWS of reaching the AMOC EDGE state. 

The network signals obtained from ocean temperature at depth (300-1000 m) have significantly larger background noise than the signals obtained from the surface temperature field (Figs.~\ref{fig:50yrPlasimDepth}, \ref{fig:GISSdepth}). Nonetheless, a weaker signal with similarities to the surface signal is visible. When comparing the results at depth between the two models, the signal is clearer in the GISS simulations. This may be related to the difference in the vertical resolution of the models (40 against 22 layers on a stretched vertical grid), leading to a much coarser representation of the mid-depth ocean in PlaSim-LSG.

We note that the relatively pronounced near-surface signal under transient forcing (Figs. \ref{PLASIMsurface}, \ref{GISSsurface}) opposes the situation under equilibrium conditions, where the network signal is stronger at depth (Fig. \ref{fig:EDGE-weak-strong-regions}). This could be due to atmospheric internal variability masking long-range ocean correlations in equilibrium conditions, while under transient forcing the coupled ocean-atmosphere system may respond in a more coherent manner. Another possible explanation is that slow degrees of freedom, such as deep-sea temperature, cannot respond as quickly to the forcing changes, such that changes in the network structure are more directly apparent in the faster upper-ocean variables. Indeed, forced transitions in multiscale systems may pass through the edge state in fast variables while bypassing it in slow variables \cite{borner_saddle_2024}.

\section{Conclusion}

Understanding the properties of the edge state separating competing stable AMOC states in a multistable regime is key to advancing our knowledge of the AMOC tipping dynamics under transient forcing, and towards developing reliable early warning signals. When the AMOC is subjected to destabilizing forcings, such as freshwater flux forcing and/or atmospheric CO$_2$ increase, the system may come in close proximity to the edge state, regardless of whether the AMOC will subsequently collapse or eventually recover. 

The determination of the edge state via edge tracking is computationally challenging and currently still prohibitively expensive in state-of-the-art earth system models, so it is tempting to try to fingerprint it using reduced order methods. Since the edge state is an unstable chaotic saddle living on the boundary separating the basins of attraction of the strong and weak AMOC state, it is reasonable to expect that one might identify qualitative differences in its fundamental dynamical properties compared to the stable attractors. 

A characteristic signature of the edge state clearly emerges when one constructs climate networks, which provide a reduced representation of both the spatial and temporal dynamics of a complex system. The networks are constructed based on instantaneous (zero-lag) correlations between the annual-mean timeseries of pairs of model grid points, meaning that they capture how ocean locations co-vary on annual timescales (rather than advective timescales). The AMOC edge state has a distinct spatial correlation pattern in the horizontal ocean temperature field at the surface and at depth, characterized by strongly enhanced teleconnections within the North Atlantic as well as emergent teleconnections with the South Atlantic. These teleconnections with the Southern Hemisphere also occur in the 21st century under the SSP2-4.5 scenario both in the PlaSim-LSG and GISS models. In PlaSim-LSG, the trajectories are known to reach the region of the chaotic edge state. The fact that a very similar network signal is detected in the transient simulations in the more complex GISS model strengthens the hypothesis that also the GISS trajectories interact with an edge state. This provides a dynamical explanation of the ensemble splitting observed in PlaSim-LSG \cite{Borner2025}, GISS \cite{GISS_Romanou} and other models \cite{oh2025noise}.

The network-based analysis presented here can be carried out on CMIP6 models showing an AMOC weakening or collapse to investigate whether an edge state is present. Future developments include assessing the performance of this spatial signature of the edge state in climate networks to serve as an early warning signal of AMOC destabilization. Further work could also explore the physical mechanisms underlying the correlation patterns captured by the network analysis.

% If you have acknowledgments, this puts in the proper section head.
\section*{Acknowledgments}
LM and MB acknowledge the financial support from the Swiss National Science Foundation (Sinergia Project No.~CRSII5\_213539). This is ClimTip contribution \#XXX; RB, VL and HD have received funding from the European Union's Horizon Europe research and innovation programme under the ClimTip project (Grant agreement No. 101137601). VL further acknowledges the partial support provided by the Horizon Europe Projects Past2Future (Grant No. 101184070) and the ARIA SCOP-PR01-P003—Advancing Tipping Point Early Warning AdvanTip project, by the European Space Agency Project PREDICT (Contract 4000146344/24/I-LR), and by the NNSFC  International Collaboration Fund for Creative Research Teams (Grant No. W2541005).

\section*{Data Availability Statement}
The code used for the network analysis and to produce figures can be found on Zenodo under \cite{Moinat2026}. PlaSim-LSG simulation output is archived by \cite{Borner2025} on Zenodo under the DOIs \href{https://doi.org/10.5281/zenodo.17053348}{10.5281/zenodo.17053348} and \href{https://doi.org/10.5281/zenodo.20270589}{10.5281/zenodo.20270589}. Due to the large amount of data, the adjacency matrix of the climate networks and additional data can be requested on demand from LM.

\section*{Author contribution statement}
LM and RB conceptualized the work with support from the other authors. RB ran the PlaSim-LSG simulations. LM performed the network analyses. LM wrote the first version of the manuscript with support from MB. All the authors reviewed and contributed to  the final version 
of the manuscript.

\section{Appendix}

Additional figures, mentioned in the main text, are shown in this Appendix.

\setcounter{figure}{0}
\renewcommand{\thefigure}{A\arabic{figure}}

\begin{figure}
\centering
\includegraphics[width=0.7\textwidth, keepaspectratio]{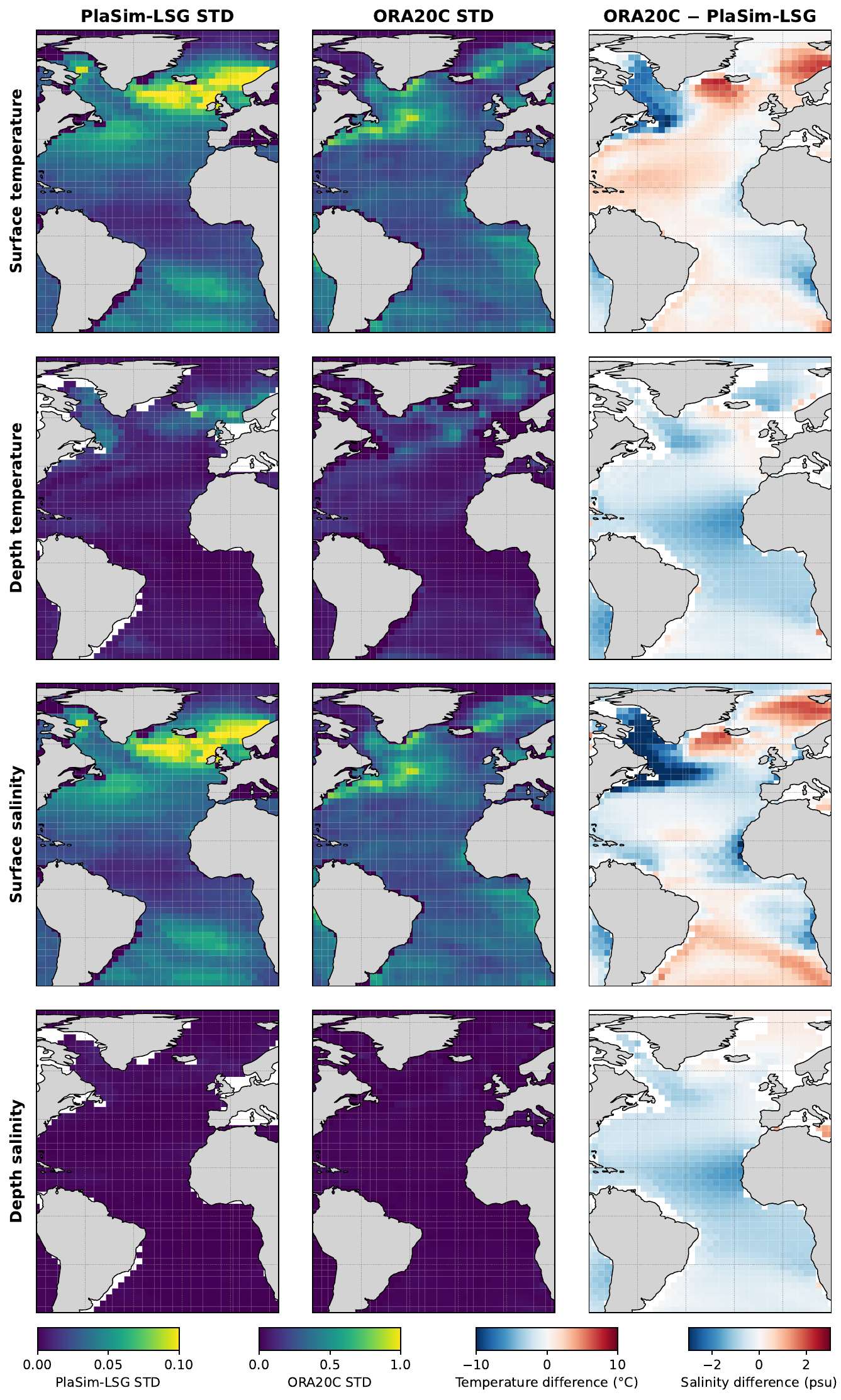}
\caption{Raw field standard deviation (STD) for PlaSim-LSG (first column), ORA20C (second column) and the difference between the ORA20C and PlaSIm-LSG raw fields (third column), over a 25 years period. This is shown for the surface temperature in °C (first row), depth temperature in °C (second row), surface salinity in psu (third row) and depth salinity in psu (fourth row).}
\label{fig:on_validation_ORA20c}
\end{figure} 

\begin{figure}  
\centering
\includegraphics[width=1.0\textwidth, keepaspectratio]{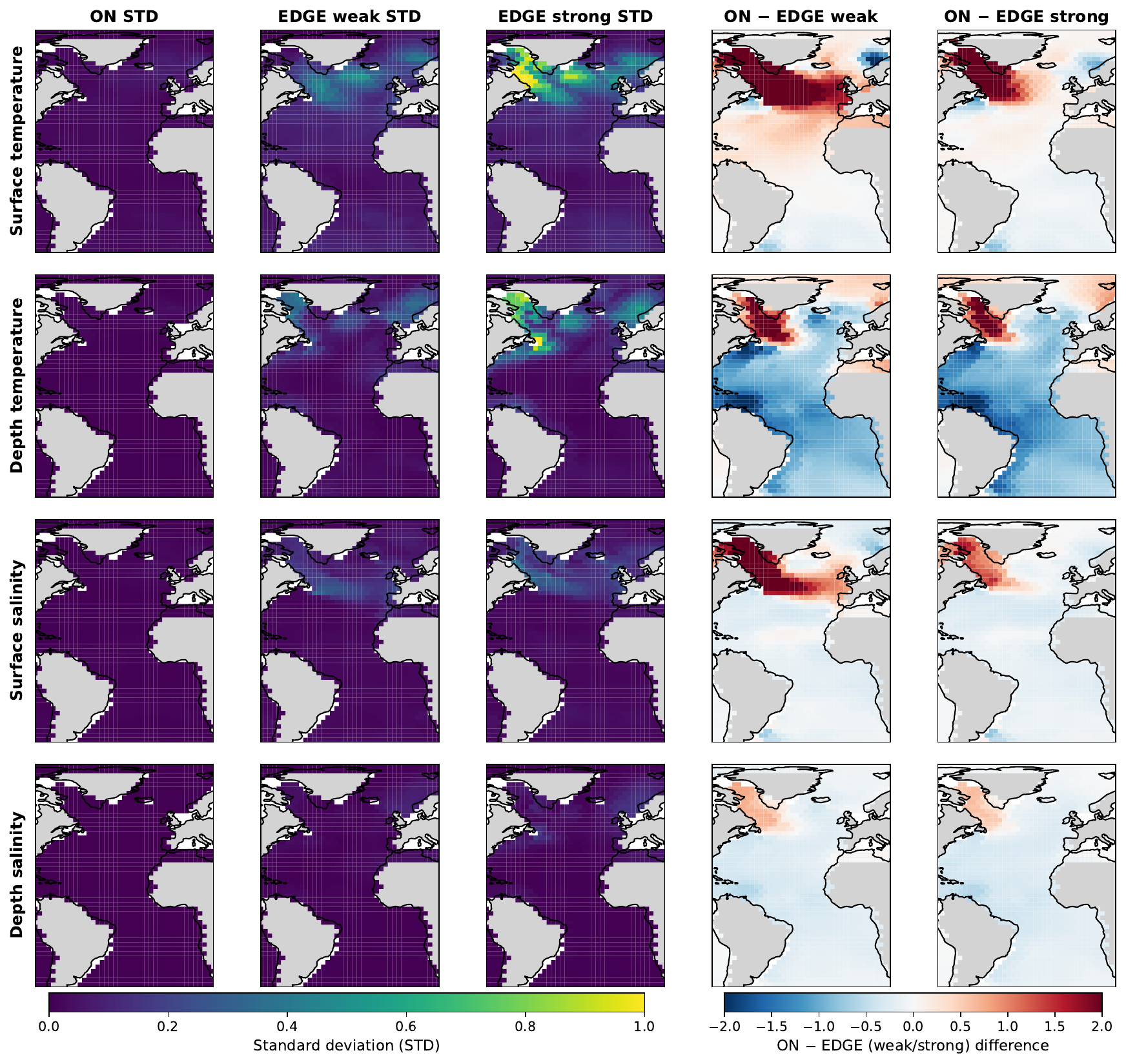}
\caption{Raw field of the standard deviation (STD) for the ON (first column), EDGE weak (second column), EDGE strong (third column) states obtained with PlaSim-LSG. Differences in the raw fields between the ON and EDGE weak (fourth column) and EDGE strong (fifth column), respectively. This is shown for the surface temperature in °C (first row), depth temperature in °C (second row), surface salinity in psu (third row) and depth salinity in psu (fourth row). }
\label{fig:differenceT}
\end{figure} 

\begin{figure}  
\centering
\includegraphics[width=1.0\textwidth, keepaspectratio]{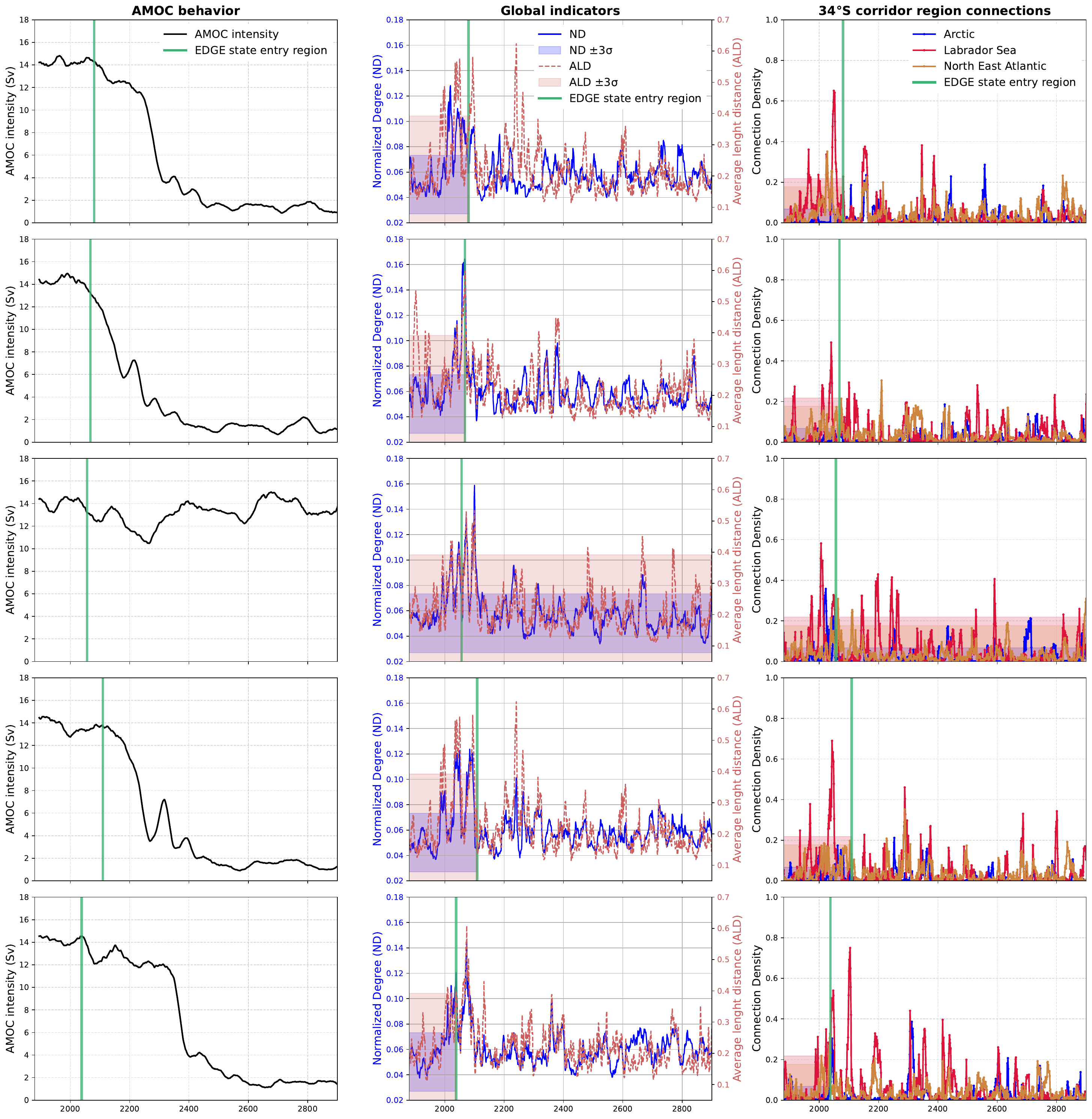}
\caption{Same as Fig.~\ref{PLASIMsurface} in five PlaSim-LSG simulation ensembles with a moving time window of 25 years.}
\label{fig:amocbehaviour25yrs}
\end{figure}

\begin{figure}  
\centering
\includegraphics[width=1.0\textwidth, keepaspectratio]{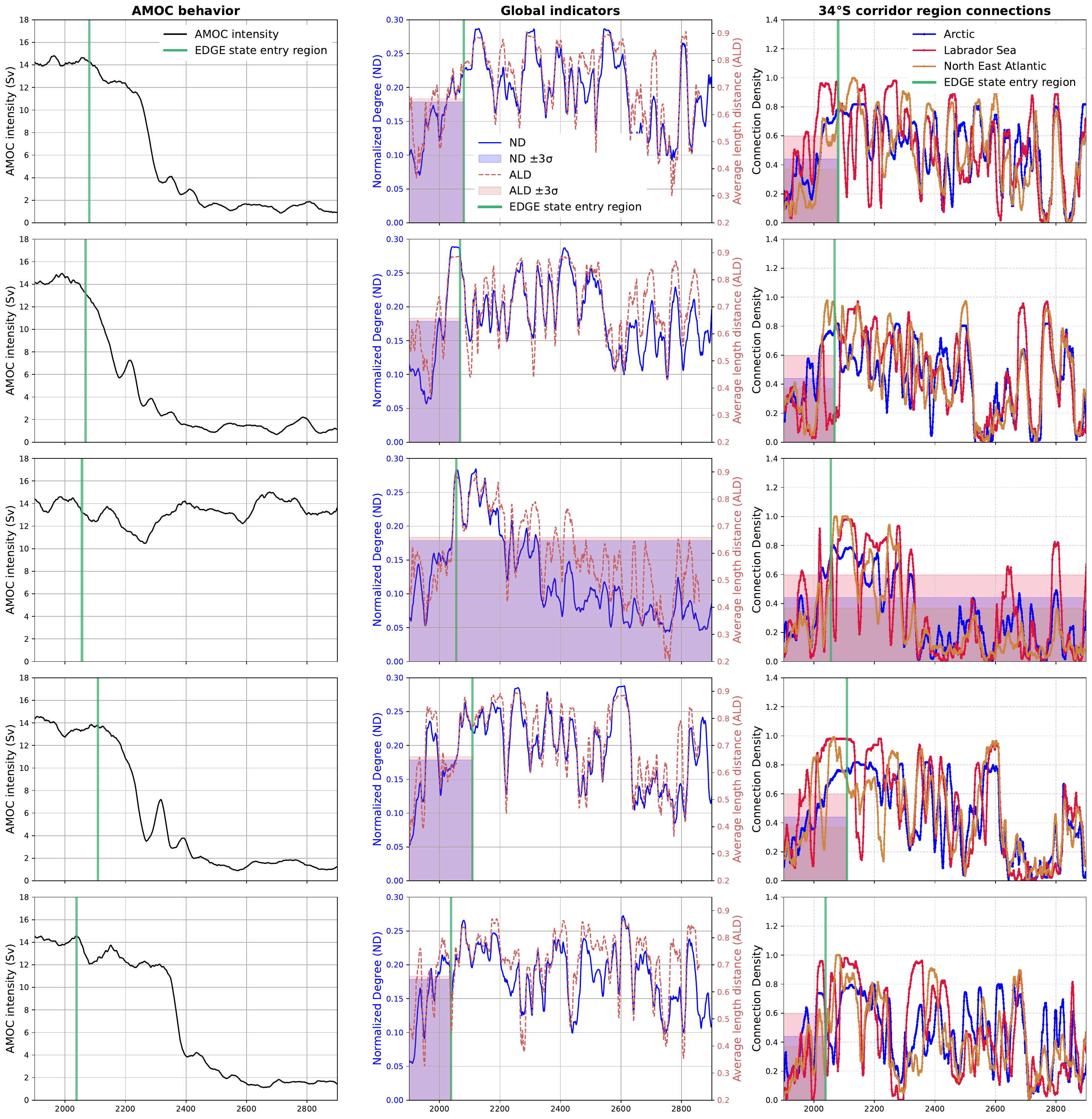}
\caption{Same as Fig.~\ref{PLASIMsurface} in five PlaSim-LSG simulation ensembles with a moving time window of 50 years at depth.}
\label{fig:50yrPlasimDepth}
\end{figure} 

\begin{figure}  
\centering
\includegraphics[width=1.0\textwidth, keepaspectratio]{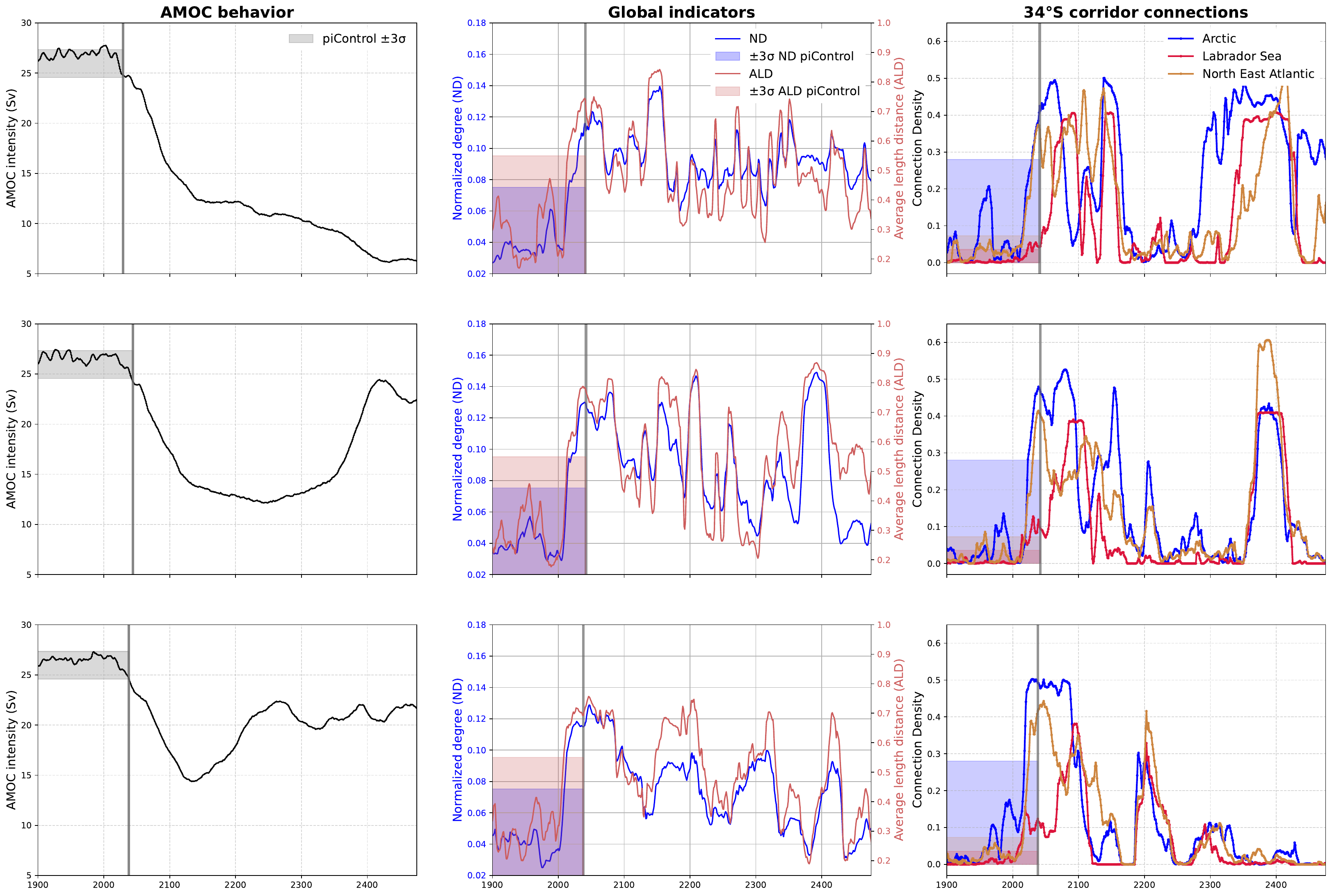}
\caption{Same as Fig.~\ref{GISSsurface} in three NASA-GISS simulations ensembles with a moving time window of 50 years at depth.}
\label{fig:GISSdepth}
\end{figure}

\clearpage

% Create the reference section using BibTeX:
\bibliography{biblioLaure}

\end{document}